\documentclass[11pt]{article}
\pdfoutput=1

\usepackage{amssymb,amsmath, graphicx}
\usepackage{latexsym}
\usepackage{mathrsfs}
\usepackage{hyperref} 
\usepackage{verbatim}
\usepackage{bm}
\usepackage{cite}
\usepackage{mathtools}
\usepackage{setspace}

\def\bea{\begin{eqnarray}}
\def\eea{\end{eqnarray}}
\def\be{\begin{equation}}
\def\ee{\end{equation}}
\def\ba{\begin{array}}
\def\ea{\end{array}}

\newcommand{\calN}{{\cal N}}

\newcommand{\calR}{{\cal R}}
\newcommand{\calD}{{\cal D}}
\newcommand{\calO}{{\cal O}}
\newcommand{\calW}{{\cal W}}
\newcommand{\calK}{{\cal K}}
\newcommand{\calL}{{\cal L}}

\newcommand{\Mpl}{{M_{\mathrm{Pl}}^2}}
\newcommand{\Mapl}{{M_{\mathrm{Pl}}}}
\newcommand{\mpl}{{M_{\mathrm{Pl}}}}

\begin{document}
\setlength\arraycolsep{2pt}

\renewcommand{\theequation}{\arabic{section}.\arabic{equation}}
\setcounter{page}{1}

\begin{titlepage}

\begin{center}

\vskip 1.0 cm

{\LARGE  \bf  Non-canonical  inflation coupled to matter}

\vskip 1.0cm

{\large
Sebasti\'an C\'espedes\footnote{E-mail:\ scespedes@damtp.cam.ac.uk} and  Anne-Christine Davis
}

\vskip 0.5cm

{\it
Department of Applied Mathematics and Theoretical Physics\\
Centre for Mathematical Sciences, University of Cambridge\\
\mbox{Wilberforce Road, Cambridge, CB3 0WA, UK.}
}
\vskip 1.5cm

\end{center}

\begin{abstract}

We compute corrections to the inflationary potential due to conformally coupled non-relativistic matter. We find  that under certain conditions of the matter coupling, inflation may be interrupted abruptly. We display this  in the superconformal Starobinsky model, where  matter is conformally coupled to the Einstein frame metric. These corrections may easily stop inflation provided that there is an initial  density of non-relativistic matter. Since these additional heavy degrees of freedom generically occur in higher dimension theories, for example as Kaluza-Klein modes,  this effect can arise in multiple scenarios. 

\end{abstract}
\end{titlepage}
\newpage
\tableofcontents

\section{Introduction} \label{intro}
\setcounter{equation}{0}
\DeclarePairedDelimiter\abs{\lvert}{\rvert}%
\DeclarePairedDelimiter\norm{\lVert}{\rVert}
The advent of  precision observations  have posed new questions on inflationary theories, with knowledge of the microphysics of inflation more necessary. Future cosmological observations, such as B-modes probe~\cite{Benson:2014qhw,Keck} or LSS surveys \cite{Alvarez:2014vva} may provide a window into the scale of inflation and henceforth an opportunity to explore phenomena occurring at  high energy scales. On the other hand some aspects of models coming from UV physics are particularly interesting today as  they may be constrained using collider physics ~\cite{Bezrukov:2007ep}. In order to study this phenomena, one should be able to  describe its characteristic manifestations in a model independent way~\cite{Cheung:2007st}.

This bottom up approach could be used to constrain string theory or 
supergravity. However, to extract inflationary viable theories from UV completion such as string theory or supergravity a large number of scalar fields  may  appear ~\cite{Baumann:2014nda}.  For example in string theory when compactifying a higher dimensional theory many scalar fields come from either Kaluza-Klein or Calabi-Yau mechanisms. If their masses are comparable to the inflaton they may intefere with the dynamics of inflation, but usually these fields are assumed to be heavy compared to the scale of inflation $H^4$,  so they are effectively integrated out from the dynamics. However, recent work has pointed out that these heavy fields can  still lead to important changes in inflation. For example, by producing corrections to the potential such as in the $\eta$ problem, or by making the inflaton speed slower~\cite{Baumann:2011su,Achucarro:2010da,Avgoustidis:2012yc,Achucarro:2012yr}. All these mechanisms are useful in model building because they may arise as  phenomenological implications of supergravity  or string theory when trying to produce inflationary potentials ~\cite{Baumann:2011nk}. Thus one can use these probes to   constrain these models . For example, if supersymmetry is a real symmetry then it has to be broken, hence inflation  could in principle be used to constrain possible mechanisms of supersymmetry
breaking ~\cite{Lyth:1998xn}. 

In this paper we present a new issue that may  arise in inflationary theories when non-relativistic matter is present in a class of non-minimally coupled 
models. Since matter is coupled via the metric we can study the effect of 
heavy degrees of freedom on inflation. A large class of theories of inflation 
are naturally written into  non-canonical framework, for example, some 
supergravity models. Moreover, Jordan frame actions arise in string theory or Kaluza Klein theories due to the existence of a dilaton field~\cite{Damour:2002mi}. 

The Starobinsky model ~\cite{Starobinsky:1980te,Whitt:1984pd}, is particularly appealing when one tries to study this kind of effect. This is because it is one of the simplest frameworks available to study inflation  whilst  remaining on solid grounds~\cite{Ade:2015lrj,Martin:2013tda}. Furthermore, there has been a resurgence of interest
in this model both because it fits the recent Planck data very well, and also because of its remarkable flexibility when embedded into supergravity ~\cite{Cecotti:1987sa,Cecotti:1987qe,Ketov:2010qz,Ellis:2013xoa,Kallosh:2013lkr,Buchmuller:2013zfa,Farakos:2013cqa,Ellis:2013nxa}. This is due to the fact that the scalar field theory can be recast as an $R^2$ correction to the Einstein-Hilbert action, without having to specify a potential resulting in a scalar gravity sector containing $R+R^2$ terms.  Moreover it has been shown that it is an attractor point in the space of single field non canonical inflation and thus allows one to study, in principle, theories with different origins\cite{Kallosh:2013maa,Kallosh:2013daa,Kallosh:2013tua,Mosk:2014cba}. One important step in a supergravity embedding was done in ~\cite{Cecotti:1987sa,Cecotti:1987qe} where, using the old and new formulation (see ~\cite{Freedman:2012zz} for an introduction),  the bosonic sector of the supergravity action was shown to be equivalent to the Starobinsky model. Nevertheless, this first approach was difficult to reconcile with a realistic supersymmetry breaking model. 
Recent advances in supersymmetry allow us to consider different realisations of 
supersymmetry consistent with a low energy inflationary limit. One succesful approach is based on a generalisation of  superconformal theories. Superconformal theories have more symmetries than the usual supersymetries, importantly an $U(1)_R$ and conformal invariance, thus the original action is more constrained. By breaking the conformal invariance one can  recover the usual  Poincare supergravity. Due to the extra symmeries the realisation of inflationary models is, in some sense, simplified. This framework can lead to inflation   written in  Jordan frame, and it was first applied to  build a supergravity model of Higgs inflation\cite{Ferrara:2010yw,Ferrara:2010in}. Furthermore, these models can be generalised to include other classes of models and, by requiring a shift symmetry in the Kahler potential, one can avoid the $\eta$ problem, Thus, they are particulary appealing for studying how phenomenological implications might affect inflation. (See \cite{deRham:2014wfa} for a discussion of why quantum corrections are not large in this case)

On the other hand one would like to study effects of modified gravity, such as the correction imposed on the Starobinsky model due to the $R^2$ term.  To do so in a model independent way many authors have studied the phenomenological implications that a modification of gravity may have~\cite{Joyce:2014kja}. It has been found that a screening mechanism has to exist in order to agree with solar system tests. One of these is the symmetron mechanism ~\cite{Hinterbichler:2010es}. Its idea is that matter which is conformally coupled to a metric in Einstein frame may induce a symmetry breaking potential when the matter density is high enough, whereas the symmetry is restored once the matter has flushed away due to cosmological evolution.
 
One interesting possibility  is to use this mechanism  within  an inflationary context. Indeed, many of the inflationary models arise as modified gravity  theories in which the coupling to gravity is  non-minimal. In this sense, Brax and Davis showed~\cite{Brax:2014baa} that for certain conformally invariant models of inflation, the equations of motion will be dependent on the density of the non-relativistic matter. Since the original action is in Jordan frame due to the conformal invariance, matter will be conformally coupled once translated to Einstein frame. Then, in a similar way to the original symmetron mechanism, for higher matter densities the inflaton field acquires a vev and  decays into one  of the minima of the potential, whereas when the matter density decays due to the inflationary expansion, the old minima will be uplifted and the inflaton will start to roll down into the new minima, thus producing inflation. Moreover the authors showed that this mechanism is more general and may be realised by a correct  use of the Damour Polyakov mechanism ~\cite{Damour:1994zq}. This idea provides a natural mechanism to set the initial conditions for inflation and was applied to
Higgs-Dilaton inflation~\cite{Brax:2014baa}.
 
 In this paper we investigate whether this mechanism can arise more generally,
in particular we will focus on the superconformal supergravity models.  To do so we will work with Jordan frame scalar gravity sectors of supergravity actions. This simplification allows us to study how massive scalar fields affect the inflationary potential because they will remain conformally coupled to the Einstein action once they are integrated out.  Scalar fields whose masses are much higher than the scale of inflation are thought to be decoupled, unless they induce a non-trivial geometry in the target space. On the contrary, we find that they may affect the dynamics of inflaton by inducing large corrections to the potential.   This result is based on the symmetron mechanism as described before. We will begin by considering the Starobinsky model, then move on to other models.
 
 In scalar tensor theories the equation of motion is modified in the presence of coupled matter as follows,
 \bea
\square\phi-V_{,\phi}-\frac{\Omega_{,\phi}}{\Omega^2}\rho=0.
\eea
where $\Omega^2(\phi)$ is the coupling between the frames $g_{\mu\nu}^E=g_{\mu\nu}^J/\Omega^2$, and $\rho$ the matter density in Einstein frame. If $\Omega^2(\phi)$  decreases exponentially, then the potential may become too steep to produce inflation for high matter density. Although the matter decays due to cosmic expansion, we found that in certain cases  the decay occurs  in a few efolds and therefore inflaton will not be able to produce adiabatic perturbations for enough time.  We will show this behaviour arises from  supergravity when a $SO(1,1)$ symmetry is used to  build a realisation of the Starobinsky model \cite{Kallosh:2013lkr} . In this model the coupling $ \Omega^2(\phi)$ arises naturally due to the conformal symmetry.   We show that when couplings of the form, 
 \bea
 \Omega^2(\phi)=1-\phi^2/6
 \eea
are considered, the presence of non-relativistic matter can affect the inflationary potential in such a way  that it becomes too steep to produce inflation. Indeed for high enough densities, the characteristic plateau of the Starobinsky model will be diminished by an exponential term. This makes the inflaton decay in a few efolds to the bottom of the potential. Henceforth  once the matter has vanished the inflaton remains stationary and unable to produce inflation.   We will show that  this result holds when one considers different types of couplings to the Ricci scalar, and that it is thus dependent on the geometry of the field space. For example,  analysing the $\alpha$ models of inflation ~\cite{Kallosh:2013yoa}, in which $\alpha$ parameterises the curvature of the Kahler manifold, one finds that for certain values of $\alpha$ inflationary evolution may be spoiled, but for values of $\alpha$ which mimic chaotic inflation the effect we described naturally disappears. We prove that this also works for other cases such as a modification of D-term inflation~\cite{Buchmuller:2013zfa}. Another interesting example is the universal attractor, where depending on the coupling function one can have  either  mimic chaotic inflation or the Starobinsky model.  This further generalises our results. We explain how  the effect we have described arises in this context. In particular for  the weak coupling regime $\abs{\xi}\ll1$ and, $ \Omega^2(\phi)=1+\xi\phi^2$, one can find a mechanism to set the initial conditions for inflation if $\xi>0$ or to spoil it if $\xi$ is negative. Moreover for the strong coupling regime $\abs{\xi}\gg 1$  the potential resembles the Starobinsky case \cite{Kallosh:2013tua}, but  the presence of matter induces a different minima and then reheating has to be more carefully considered.

 This article will be structured as follows. In section two we analyse non-canonical models of inflation and introduce some of the main tools  needed to proceed. We also give details of the Starobinsky model and possible embeddings in supergravity.  We then discuss superconformal $\alpha$-attractors, which generalise the Starobinsky model.
 In  section three we analyse how matter is coupled to inflation and its affect on the inflaton potential via the symmetron mechanism. We explain under which circumstances inflation is spoiled by the presence of non-relativistic matter and we described how matter changes the attractor behaviour of single field inflation. By generalising to the universal attractor models, we described how variations in the coupling may produce different implications.  We  analyse possible issues that may affect our conclusions. Finally we conclude by giving some possible directions for further work.


\section{Non-canonical inflation} \label{intro}
\setcounter{equation}{0}
In this section we first review the standard formalism used when studying non-canonical inflationary theories. We will then focus on explaining  in some detail the  Starobinsky model of inflation, and give some examples for supergravity theories in which the Starobinsky model can be embedded.

 By making some assumptions we will simplify one particular superconformal supergravity which  can give rise to an $SO(1,1)$ invariant scalar tensor theory. 
We will show how this action is equivalent to the Starobinsky model. Finally,   we extend this framework to include other more general types  of models of inflation, including chaotic inflation.

\subsection{Basic setup}
In this paper we work with non-canonical models of inflation with action given by,

 \bea
 S=\int d^4x\sqrt{-g} \left\{\frac{1}{2}\Omega^2(\phi) R-\frac{K(\phi)}{2}\partial_\mu\phi\partial^\mu\phi - V_J(\phi)\right\}, 
\label{action:NCInflation}
 \eea
 where $\Omega(\phi)$ is the non-minimal coupling with the Ricci scalar and $K(\phi)$ is the non-canonical kinetic term. This action is very general. For example, with coupling function  of the form $\Omega=1+\xi\phi^2$ inflation can be 
realised using the Higgs boson. Furthermore, requiring the action to be conformally invariant puts constraints on the coupling functions. For  Weyl invariance, i.e. scale invariance,  one  has that $\Omega^2(\phi)=-6K(\phi)$. 
 
 The coupling function $\Omega(\phi)$ can be made constant by performing a conformal transformation,
  \bea
 \phi\rightarrow\Omega(x)\phi, \hspace{1cm} g_{\mu\nu}\rightarrow\frac{1}{\Omega(x)^2}g_{\mu\nu}.
 \eea
 When $\Omega(\phi)=\Mpl$, the action is in the Einstein frame, whereas when $\Omega(\phi)$ is not a constant, it is in Jordan frame. In general, this will introduce new terms in the kinetic energy because of the transformation law for the Ricci scalar. The action in Einstein frame is,
 
 \bea
 S=\int d^4x\sqrt{-g} \left\{\frac{1}{2}R-\frac{1}{2}\left\{\frac{K(\phi)}{\Omega^2}+6(\log\Omega(\phi))'^2\right\}\partial_\mu\phi\partial^\mu\phi - V_E(\phi)\right\}, 
\label{action:NCInflationEinstein}
 \eea
  where $V_E=V_J/\Omega^4$. Although it is preferable to change to Einstein frame when working with inflation because observables are simpler to work with, sometimes working in the Jordan frame is useful because it makes physics clearer. 
Both frames should make the same predictions because they just correspond to different coordinate systems, a fact that we exploit during the course of this 
paper. Finally, note that the kinetic coupling  can be made canonical by a field redefinition, provided that $K$ is well defined and non-zero in the domain of interest.

\subsection{Starobinsky model}
One of the first models of inflation was based on  a modification of the Einstein Hilbert action proposed by Starobinsky~\cite{Starobinsky:1980te}. In this framework the new action contains a higher derivative term which is minimally coupled to the metric.  Its action is  given by
\bea
 S_S=\frac{\Mpl}{2}\int d^4x\sqrt{-g} \left(R+\frac{1}{6\mpl M^2}R^2\right),
 \label{action:Starobinsky1}
 \eea
where there is a new coupling constant $M^2$ which may  be constrained, for example, using inflation in which case $M^2\ll \Mpl$ . 
To see more clearly how this model can produce an inflationary epoch let us rewrite it  as a scalar tensor theory introducing an auxiliary field $\psi$. The action is,
\bea
 S_S=\int d^4x\sqrt{-g} \left(\frac{\Mpl}{2}R+\frac{1}{M}R\psi-3\psi^2\right).
 \label{action:Starobinsky2}
\eea
Since $\psi$ is just a lagrange multiplier, the action can be transformed into the original form (\ref{action:Starobinsky1}) by solving the  constraint equation for the field $\psi$. 

To count the degrees of freedom of the theory let us rewrite the action ({\ref{action:Starobinsky2}) in Einstein frame by means of the transformation,
\bea
g_{\mu\nu}\rightarrow e^{-\sqrt{2/3}\phi/\mpl} g_{\mu\nu}=\left(1+\frac{2\psi}{M\mpl}\right)^{-1}g_{\mu\nu}
\eea
which leads to the more familiar scalar field action,
\bea
 S=\int d^4x\sqrt{-g^E} \left[\frac{\Mpl}{2}R-\frac{1}{2}\partial_\mu\phi\partial^\mu\phi-\frac{3}{4}M^2\mpl^4\left(1-e^{-\sqrt{2/3}\frac{\phi}{\mpl}},\right)^2\right].
\label{action:inflStarobinsky}
\eea
 Therefore, we see that the effect of adding a $R^2$ term in (\ref{action:Starobinsky1}), is to introduce a scalar degree of freedom in addition to the two graviton helicities already present in the Hilbert-Einstein action. Now we have the basic ingredients for single field inflation. 
 
 The potential in (\ref{action:inflStarobinsky}) grows exponentially for negative $\phi$ whereas it has a plateau for positive $\phi$. Hence it  is  possible to have inflation if the field starts rolling down from right of the potential. Indeed since inflation will terminate when $\epsilon=1$, we have that there is a critical value $\phi_c$ for which a field starting to roll down at $\phi_0>\phi_c$, can produce enough efolds $N$  of inflation, as required by CMB observations. This issue will be considered later in this work.
 
 Moreover, one can predict the value for the tilt and the tensor to scalar ratio solely  in terms of $N$, which are,
\bea
n_s-1\approx -\frac{2}{N}, \hspace{1cm}r\approx\frac{12}{N^2},
\eea
Thus we have that for $N\approx 60$, $n_s\sim 0.03$ and $r\approx \sim 0.004$ which are compatible with the Planck data.
The recent interest in the Starobinsky model is because it seems to be favoured by recent cosmological experiments. This has motivated further theoretical research. For example, it has been pointed out that Higgs inflation~\cite{Bezrukov:2007ep} and the Starobinsky model are equivalent for a  particular limit of the action (\ref{action:NCInflationEinstein})\cite{Kehagias:2013mya}. Moreover, it was suggested that they form part of a wider family of models, all of the which have similar predictions for inflation, called universal attractors~\cite{Kallosh:2013maa,Kallosh:2013hoa,Kallosh:2013daa}. 


\subsection{Inflation from superconformal gravity}
The canonical superconformal supergravity (CSS) approach developed in \cite{Kallosh:2013daa,Ferrara:2010yw}, is particularly  useful for model building because it can provide a supergravity context to Jordan frame actions.  We will now review part of the framework relevant to our work.
Superconformal symmetry has for a long time been used as a method to build supergravity multiplets.  Here the  action is  invariant under superconformal transformations. This transformations includes,  conformal transformations and $U(1)_R$ transformations. Therefore one has to fix the gauge to recover the standard supergravity formulation. Due to a higher degree of symmetry the superconformal action is much simpler than  Poincare supergravity and thus more suitable for phenomenological applications.  
The action  composed of $X^I$ bosonic fields, $\Omega^I$ fermionic fields, and $F^I$ auxiliary fields is given by
 \bea
\calL=\calN(X,\bar{X})\vert_D+\calW(X)\vert_F+f_{\alpha\beta}\bar\lambda^\alpha_L\lambda^\beta_L\vert_F.
\label{action:SG}
 \eea 
The function $\calN(X,\bar{X})$ is related to the Kahler potential and $\calW(X)\vert_F$ to the super potential. The last part is related to the kinetic term for the vector fields and will not be important for us.  The subindex F and D denotes the  last term of the multiplet. Note that the action is written in direct analogy with the supersymmetric Wess Zumino model. 
 The conformal weights of each function are in order $2,\ 3,\ 0$.   One of the multiplets $X^0,\ \Omega^0\ F^0$, can be viewed as a compensator multiplet, i.e.. fields who are fixed to break the conformal symmetry. 
 The scalar fields form a Kahler manifold whose metric is given by
 \bea
 G_{I\bar J}=\frac{\partial\calN(X,\bar{X})}{\partial X^I\partial \bar X^{\bar{J}}}.
 \eea
 We will now describe the the $SU(2,2\vert 1)$ superconformal model developed in~\cite{Ferrara:2010yw,Ferrara:2010in}. 

\subsubsection{Canonical superconformal supergravity}
\label{CSS}
Usually superconformal symmetry is fixed at very early stages because it is used to formulate Poincare supergravity. However, in this formalism the action has 
a very simple bosonic sector, it can be written in Jordan frame; thus it is suitable when studying non-canonical models of inflation. In any case, the action (\ref{action:SG}) is still very complicated and to simplify let  us make two assumptions.
  First, we make a choice of Kahler manifold invariant under $SU(1,N)$,
\bea
\calN(X,\bar{X})=-\abs{X^0}^2+\abs{X^{\alpha}}^2 \hspace{1cm} \alpha=1,\ldots,N.
\eea
This gives a flat Kahler manifold metric, given by
\bea
G_{I\bar{J}}=\partial_I\partial_{\bar{J}}\calN=\eta_{I\bar{J}}.
\eea
Also we choose a general  cubic superpotential independent of $X^0$, 
\bea
\calW(X)&=&\frac{1}{3}d_{\alpha\beta\gamma}X^\alpha X^\beta X^\gamma,\\
\calW_0&=&0,\nonumber
\eea
where $d_{\alpha\beta\gamma}$ is a numerical matrix. The scalar gravity part of the action becomes, 
\bea
\frac{1}{\sqrt{-g}}\calL_{SG}=-\frac{1}{6}\calN(X,\bar{X})R-G^{I\bar{J}}\calD_\mu X^I\calD_\mu\bar{X}^J-G^{I\bar{J}}\calW_I\calW_{\bar{J}}, \hspace{1cm} I, \bar{I}=0, 1,...,n.
\label{action:sugra}
\eea
We still need to fix the gauge by a change of variables from the basis $\{X^I\}$ to a basis $\{y,\ z^\alpha\}$, using $X^I=yZ^I(z)$.
The dilaton and $U(1)$ symmetry are fixed by choosing, $\calN(X,\bar{X})=-\abs{X^0}^2+\abs{X^{\alpha}}^2=\Phi(z,\bar z)$, and 
\bea
X^0=\bar X^{\bar 0}=\sqrt{3}\mpl, \hspace{0.5cm} y=\bar y=1,\hspace{0.5cm} X^{\alpha}=z^\alpha.
\eea
This choice results in the compensator fields decoupling from the "physical" fields, thus the functions become
\bea
\hat\Phi(z,\bar z)=-3\Mpl+\delta_{\alpha\bar\beta}z^\alpha z^{\bar\beta}, \hspace{1cm} \calW(X)=W(z)=\frac{1}{3}d_{\alpha\beta\gamma}z^\alpha z^\beta z^\gamma,
\eea
To recover the usual poincare supergravity one has to fix the compensator fields. For example by choosing $X^0=\bar X^{\bar 0}=\sqrt{3}$ we get,
\bea
\calN(X,\bar X)\vert_{X^0=\bar{X}^{\bar{0}}=\sqrt{3}}&=&-3e^{\frac{1}{3}\calK(X^I,\bar X^{\bar I})},\\
\label{set:Kahler}
\calW(X^I)\vert_{X^0=\sqrt{3}}&=&W(X^I).
\eea
This formalism was used to build a supergravity generalisation of  Higgs inflation~\cite{Ferrara:2010in}, but has been applied to several other models~\cite{Kallosh:2010xz,Buchmuller:2012ex,Ferrara:2013rsa,Kallosh:2013pby}. What it makes it particularly interesting is the fact that it is simple to set up viable models of inflation in which the fields other the inflaton are rendered stable.
 
\subsubsection{Starobinsky model}
The Starobinsky model can be embedded in supergravity. As explained before, since the scalar tensor action (\ref{action:inflStarobinsky}) is equivalent to (\ref{action:Starobinsky1}), it is not necessary to specify the potential, but it is sufficient to find a supermultiplet containing the term $R^2$. This attempt was done by  Ceccoti \textit{et al.} during the eighties~\cite{Cecotti:1987sa,Cecotti:1987qe}, where it was shown that the supermultiplet could be realised in either old or new supergravity. Whereas the original models were unstable  more accurate versions have became available. Furthermore, one can also try to build an action by specifying the potential. This possibility  was developed by Kallosh, Linde, \textit{et al.} using CSS described earlier \cite{Kallosh:2013lkr}. In this framework one starts with a superconformal theory which recovers supergravity once the conformal invariance is broken. The theory is composed of three chiral superfields, 
\bea
X^I=(X^0, X^1=\Phi, X^2=S)
\eea
where $X^0$ is a conformon field, $X^1$ is the inflaton $\Phi$ and $X^2=S$ is a Goldstino superfield. The role of $S$ is to allow an stable inflationary trajectory. Taking the functions
\bea
\calN(X,\bar X)&=&-\abs{X^0}^2 \exp\left( -\frac{\abs{S}^2}{\abs{X^0}^2}+\frac{1}{2}\left(\frac{\Phi}{X^0}-\frac{\bar{\Phi}}{\bar X^{\bar 0}}\right)^2+\zeta\frac{\abs{S}^4}{\abs{X^0}^4}\right),\\
\calW(X^0,\Phi,S)&=&\frac{M}{2\sqrt{3}}S(X^0)^2\left(1-e^{-\frac{2\Phi}{\sqrt{3}}}\right)^2.
\eea
The field $S$ has to be introduced in order to spontaneausly break the supersymmetry, which is achieved when the auxiliary field is vanishing, $F=\calW_S=0$. Following \ref{set:Kahler}, the corresponding Kahler manifold is given by,
\bea
\calK(\Phi,\bar\Phi,S,\bar S)=S\bar S-\frac{(\Phi-\bar\Phi)^2}{2}-\zeta(S\bar S)^2.
\eea

Note that there is a shift symmetry for the field $\Phi$, and thus large corrections to the mass are forbidden. To get an inflationary action after fixing the conformal symmetry one assumes that there is an inflationary trajectory along the space defined by $S=\rm{Im} \Phi=0 $. This direction becomes stable by the term proportional to  $\zeta$. Then one recovers a scalar gravity action which is equivalent to the Starobinsky model with potential $V\sim \left(1-e^{-\sqrt{2/3}\frac{\varphi}{\mpl}},\right)^2$. 
Notice that this embedding is not unique, other possibilites has been explored but using different aproaches to supergravity \cite{Ketov:2010qz,Ellis:2013xoa,Buchmuller:2013zfa}. Nevertheless the one we described has the advantage of producing an action which starts in Jordan frame. 



\subsection{Conformal inflation}
Using the previous choice of gauge one may just simply assume that the low energy inflationary limit exists and study its phenomenological consequences. In order to do so we fix the field $S=0$ but also work with the real part of the fields $X^0,\ \Phi$. Following \cite{Kallosh:2013lkr} we start by studying the  phenomenological Lagrangian given by the action,

\bea
 S=\int d^4x\sqrt{-g} \left\{\left(\frac{\chi^2}{12}-\frac{\phi^2}{12}\right)R+\frac{1}{2}\partial_\mu\chi\partial^\mu\chi-\frac{1}{2}\partial_\mu\phi\partial^\mu\phi\ - \frac{\lambda}{4}\phi^2(\phi-\chi)^2\right\},
\label{action:conformalinv}
 \eea
 which was shown to be equivalent to the Starobinsky model. This models spontaneously breaks the conformal symmetry and thus there is an extra scalar degree of freedom compared to the conformally invariant De Sitter case,
\bea
 S=\int d^4x\sqrt{-g} \left\{\left(\frac{\chi^2}{12}-\frac{\phi^2}{12}\right)R+\frac{1}{2}\partial_\mu\chi\partial^\mu\chi-\frac{1}{2}\partial_\mu\phi\partial^\mu\phi\ - \frac{\lambda}{4}(\phi^2-\chi^2)^2\right\},
\label{action:conformalinvDS}
 \eea
 which is invariant under conformal transformations,
\bea
 \phi\rightarrow\Omega(x)\phi, \hspace{1cm} g_{\mu\nu}\rightarrow\frac{1}{\Omega^2(x)}g_{\mu\nu}.
 \label{transf:conformal}
 \eea
 Furthermore (\ref{action:conformalinv})  is invariant under  an $SO(1,1)$ transformation in the fields. It is a generalisation of the conformally invariant action,
\bea
 S=\int d^4x\sqrt{-g} \left\{-\frac{\phi^2}{12}R-\frac{1}{2}\partial_\mu\phi\partial^\mu\phi \right\},
 \label{action:singlefieldconf}
\eea
 which is equivalent to Einstein-Hilbert gravity just by a conformal transformation. In the above action the kinetic term has the wrong sign but the theory is not ill defined because the $\phi$ field is unphysical. Indeed the above theory just propagates two degrees of freedom corresponding to the two graviton polarizations. This redundancy might be removed by fixing the gauge freedom choosing,
 \bea
 \phi=\sqrt{6}\Mpl
 \eea
which leads to the Einstein-Hilbert action. Moreover this construction proved to be fundamental when building supergravity actions. As we mentioned before, by requiring invariance under superconformal transformations, it is possible to build a Poincare supergravity upon gauge fixing. 

Now coming back to (\ref{action:conformalinv}), this action is well defined for all field space except for the point where $\phi=\chi$. By doing the transformation $\phi=r\sinh(\varphi/\sqrt 6),\  \phi=r\cosh(\varphi/\sqrt 6)$ the action becomes the Starobinski model,
\bea
 S=\int d^4x\sqrt{-g^E} \left\{\frac{\Mpl}{2}R-\frac{1}{2}\partial_\mu\varphi\partial^\mu\varphi-\frac{3}{4}\lambda\mpl^4\left(1-e^{-\sqrt{2/3}\frac{\varphi}{\mpl}},\right)^2\right\} ,
\eea after transforming to Einstein frame. The above condition is implicit in the fact that for going to Einstein frame, $r\neq 0$. Also, as the action does not depend on $r$,  there are three degrees of freedom in contrast to  (\ref{action:singlefieldconf}).
 
 Recently, it has been pointed out that this  conformal symmetry is a fake symmetry because its current is zero, and thus  has no net effect on the physics \cite{Hertzberg:2014aha,Jackiw:2014koa}
One then should be careful when considering properties derived from this symmetry.

 As we are mainly interested in the gauged case, for us this fact will be unimportant. For our analysis it is more appropriate to proceed by choosing the gauge,
 \bea
 \chi=\sqrt{6}\Mpl,
\label{gaugechoice:1}
 \eea
 This gauge is related to the rapidity of the fields.
 One can consider several other gauge choices, for example $\chi^2-\phi^2=6$, that lead to  equivalent Einstein frame actions for the scalar field. Nevertheless,  one has to be very careful  when generalising this theory, because some issues will appear later when coupling matter to the action (\ref{action:conformalinv}). 
The action,  in our gauge choice $\chi=\sqrt{6}\Mpl$, becomes
 \bea
 S=\int d^4x\sqrt{-g} \left\{\frac{\Mpl}{2}\left(1-\frac{\phi^2}{6\Mpl}\right) R-\frac{1}{2}\partial_\mu\phi\partial^\mu\phi\ -\frac{3}{2}\lambda\Mpl\phi^2(\frac{\phi}{\sqrt{6}\Mapl}-1)^2 \right\}.
 \eea
which is a scalar tensor theory where  the  kinetic term has the correct sign. To analyse the inflationary limit  we can transform this action into Einstein frame by rescaling the metric, using, 
 \bea
 g_{\mu\nu}^J=\frac{g_{\mu\nu}^E}{1-\frac{\phi^2}{6\Mpl}}.
 \eea
Notice that the transformation is undefined for $\frac{\phi^2}{6\Mpl}=1$, which is equivalent to our earlier observation that the action is defined for all field space except for  $\chi=\phi$.
Doing so, we get the following action,
\bea
 S=\int d^4x\sqrt{-g^E} \left[\frac{\Mpl}{2}R-\frac{1}{2}\left(\Omega(\phi)^{-1}+\frac{3}{2}[\log\Omega(\phi)^2]'\right)\partial_\mu\phi\partial^\mu\phi -\lambda\Mpl\phi^2(\frac{\phi}{\sqrt{6}\Mapl}-1)^2 \right],
\label{action:transf}
\eea
where we  defined $\Omega(\phi)\equiv 1-\frac{\phi^2}{6\Mpl}$. The logarithms come from the transformation law for the Riemann tensor. Note that the action will have, in general, a non-canonical kinetic term. Expanding the kinetic term  in (\ref{action:transf}), 
\bea
\left(\Omega(\phi)^{-1}+\frac{3}{2}[\log\Omega(\phi)^2]'\right)\partial_\mu\phi\partial^\mu\phi=\frac{1}{(1-\phi^2/6)^2}\partial_\mu\phi\partial^\mu\phi.
\eea
The field can be canonically normalised by performing a field redefinition,
\bea
\frac{d\phi}{d\varphi}=\frac{1}{1-\phi^2/6\Mpl}
\eea
This relation can be easily integrated, which leads to 
\bea
\phi=\sqrt{6}\Mapl\tanh(\frac{\varphi}{\sqrt{6}\mpl}),
\eea
where the field $\varphi$ goes from minus infinity to infinity. Now, we will have an action for a scalar field which is in Einstein frame and also has a canonical kinetic term. Replacing the new field into the action, we get 
\bea
 S=\int d^4x\sqrt{-g^E} \left\{\frac{\Mpl}{2}R-\frac{1}{2}\partial_\mu\varphi\partial^\mu\varphi-\frac{3}{4}\lambda\mpl^4\left(1-e^{-\sqrt{2/3}\frac{\varphi}{\mpl}},\right)^2\right\} ,
\eea
which is the same action for the Starobinsky model described earlier (\ref{action:inflStarobinsky}), with $\lambda=M^2$.
\subsubsection{$\alpha$ models}
The  formalism we described before can be used to generalise in order to obtain more   diverse scenarios which are different from the Starobinsky model. For example, in the $\alpha$ model, proposed by Kallosh and Linde~\cite{Kallosh:2013yoa}, the Kahler potential  is a deformation of the  simpler flat $SO(1,1)$ model we considered earlier. Its Kahler potential is given by

\bea
\calN (X_0, X_1, S)=-\abs {X_0}^2\left[ 1-\frac{\abs{X_1}^2+\abs{S}^2}{\abs{X_0}^2}\right]^\alpha, \label{Kahler}
\eea 
where we can see $\calN(X,\bar X )$  is no longer flat nor invariant under $SO(1,2)$, with $\alpha$ parametrising deviations from it. Indeed, for $\alpha=1$ we recover the original theory. Due to the new form of $\calN (X_0, X_1, S)$, we  also have a non-canonical kinetic term,
\bea
G_{I\bar{J}}=\partial_I\partial_{\bar{J}}\calN\equiv\frac{\partial\calN(X\bar{X})}{\partial X^I\partial\bar{X}^{\bar{J}}}.
\eea
The superpotential is now given by
\bea
\calW=S((X^0)^2)f(X^1/X^0)\left[1-\frac{(X^1)^2}{(X^0)^2}\right]^{(3\alpha-1)/2}.
\eea
where to admit more generality, the potential will depend on the arbitrary scalar function $f(X^1/X^0)$.
For our purpose, it will be enough to assume that the theory has an inflationary limit where the moduli fields are stabilised, therefore  the action can be written in a  form similar to   (\ref{action:conformalinv}).  We can use as a guide the same framework we studied at the beginning of this section, where $\alpha$ will parametrise deviations from the action (\ref{action:conformalinv}). Nevertheless, the action is no longer invariant under $SO(1,1)$, and we need to set $X_0=\sqrt{3}\Mpl$ to get inflation. The potential term is  then given by
\bea
V=G^{SS}\abs[\Big]{\frac{\partial \calW}{\partial S}}^2.
\eea
 To describe inflation, the bosonic action in Jordan frame is,
\bea
S(\phi)= \int d^4x\sqrt{-g}\left\{\frac{1}{2}\left(1-\frac{\phi^2}{6} \right)^\alpha R-\frac{1}{2}\frac{\alpha-\alpha^2\phi^2/6}{(1-\phi^2/6)^{2-\alpha}} (\partial\phi)^2-V(\phi) \right\},
\label{action:alphaJordan}
\eea
where the non-canonical kinetic term is due to the shape of the Kahler potential (\ref{Kahler}).  In Einstein frame (\ref{action:alphaJordan}) reduces to 
\bea
 S=\int d^4x\sqrt{-g} \left\{R-\frac{\alpha/2}{(1-\phi^2/6)^2}(\partial_\mu\phi)^2 -f^2(\phi/\sqrt{6})\right\}, 
\eea
We can canonically normalise the kinetic energy with the transformation
\bea
\phi=\tanh(\frac{\varphi}{\sqrt{6\alpha}}),
\eea
to get the action 
\bea
 S=\int d^4x\sqrt{-g} \left\{R-\frac{1}{2}(\partial_\mu\varphi)^2 -f^2(\tanh(\varphi/\sqrt{6\alpha}))\right\}, 
\label{action:alphaEinstein}
\eea
note that the potential depends on the function $\tanh(\varphi/\sqrt{6\alpha})$. It has been pointed out in \cite{Kallosh:2013daa}
\bea
\Omega=\left(\sinh\left(\frac{\varphi}{\sqrt{6}\alpha}\right)\right)^{2\alpha}.
\eea
Choosing $f(\phi/\sqrt{g})=  \frac{\lambda\phi^2}{(\phi/\sqrt{6}+1)^2}$, we can recover  Starobinsky like potentials of the form,
\bea
 V(\varphi)\sim\left(1-e^{-\sqrt{\frac{2}{3\alpha}}\varphi}\right)^2,
\eea
which is the same as that for the Starobinsky model when $\alpha=1$. Nonetheless, note that  for larger $\alpha$ we do not get exponential terms in the potential but something more similar to chaotic inflation. One can calculate the inflationary parameters in terms of  $\alpha$, which turn out to be,
\bea
n_s=1-\frac{2}{N}, \hspace{1cm} r\approx \frac{12\alpha}{N^2},
\eea
We thus see that for large $\alpha$, $r$ may be of the same magnitude as that predicted by other models such as chaotic inflation. This particular feature is very important since $r$, once detected, may be sufficiently large to rule out the Starobinsky model, but not all the $\alpha$ models. 
Finally, supersymmetry breaking places further theoretical  constraints on $\alpha$. For example, to avoid tachyonic instabilities that may arise from $S$ during an inflationary trajectory, $\alpha>1/3$.
Furthermore it was found that $\alpha$ is related to the curvature of the Kahler manifold as, 
\bea
\calR_k=-\frac{2}{\alpha},
\eea
Therefore we can identify the different models of inflation as related to a different geometry in field space. 
\section{Matter coupled to inflation}
\label{gralmodel}
\setcounter{equation}{0}
We  will now examine what happens when matter is coupled to  non-canonical inflation. We will also show how the symmetron mechanism can be used to further constrain the possible scenarios for the Starobinsky model.
In the low energy approximation several extra fields have been integrated out
and no longer participate in the dynamics. Hence, we can write the action as,
\bea
S=S_{\rm infl}+S_{\rm matter}
\label{action:coupled}
\eea
where 
\bea
S_{\rm matter}=\int d^4x\sqrt{-g^J}\calL_M(g_{\mu\nu}^J,\psi^I),
\eea
is the action for the non-relativistic integrated out matter. Kaluza-Klein 
towers would contribute to $S_M$ in a similar way. Now, assuming that a UV 
complete theory like supergravity gives an inflationary limit in Jordan frame, we can re-express the action in Einstein frame. To do so, we make a conformal transformation in the non-relativistic matter action as well. It follows that under the  (\ref{transf:conformal}), the action transforms as,
\bea
\int d^4x\sqrt{-g^J}\calL_M(g_{\mu\nu}^J,\psi^I)\rightarrow \int d^4x\frac{\sqrt{-g^E}}{\Omega(x)^4}\calL_M(\Omega^{-2}g_{\mu\nu}^E,\psi^I).
\eea
Varying the action (\ref{action:coupled}) ,with respect to inflaton $\phi$ we obtain the following   equations of motion,
\bea
\square\phi-V_{,\phi}+\Omega(\phi)^{-5}\Omega_{,\phi}T^J=0,
\label{eq:mattercoupledJordan}
\eea
where $T^J=g_J^{\mu\nu}T^J_{\mu\nu}$ is the trace of the energy momentum tensor written in Jordan frame, $T^J_{\mu\nu}=(-2/\sqrt{-g^J})\delta\calL_m/\delta g_J^{\mu\nu}$. This momentum energy tensor is covariantly conserved. Recalling that for non-relativistic matter  $T^J=-\rho^J$, we can rewrite  this equation for Einstein frame as $\rho^E=\Omega^{-3}\rho^J$. This quantity   is also  conserved in Einstein frame\cite{Joyce:2014kja}. Hence, the equation (\ref{eq:mattercoupledJordan}) becomes,
\bea
\square\phi-V_{,\phi}-\frac{\Omega_{,\phi}}{\Omega^2}\rho=0.
\eea
We can think of this equation as a   scalar field   governed by an effective potential $V_{\rm eff}(\phi)=V(\phi)-(\Omega(\phi))^{-1}\rho$, wich is  dependent on the  density of matter. Note that $\rho$ will decay as $a^{-3}$ as  the universe expands, as it  occurs  in an inflationary background. Thus it is consistent to assume  the universe  contained non-relativistic matter before inflation since all such matter would disappear as the universe starts inflating. 

However, since there is a coupling dependence the potential could still be modified  before the matter was diluted away. This mechanism was first studied by Hinterbichler and Khoury~\cite{Hinterbichler:2010es}  in the context of modified gravity. They found that non-relativistic matter could induce a breaking in the symmetry of the system which is restored once the matter fades away. 

This was applied  to inflation~\cite{Brax:2014baa,Dong:2013swa} , where it was  shown that for a density $\rho$  larger than the energy density of the inflaton, the system could be governed by a broken symmetry potential. This in turn,  could lead to a  natural mechanism to set the initial conditions for inflation, because when matter dominates the inflation decays to the bottom of the effective potential but once matter flushes away, the original inflationary potential is restored and the field naturally starts to roll down.  Whereas  in~\cite{Dong:2013swa}  the conformal coupling was introduce by hand, in~\cite{Brax:2014baa} the studied  model was conformally  invariant under $SO(N)$ from the beginning. This case was inspired by  Higgs-Dilaton inflation~\cite{GarciaBellido:2011de} where it also was applied. In these models the coupling is given by the symmetry and the mechanism works more naturally. We will now apply this mechanism  to the Starobinsky model in the context of a conformal field theory. 

For our  case, let us start from the bosonic sector of the action described in (\ref{CSS}).  For simplicity let us assume that there is an inflationary limit which has as an action (\ref{action:conformalinv}). Once we have set the gauge by fixing $\chi=\sqrt{6}\mpl$ in eq.~\ref{gaugechoice:1} the coupling is $\Omega^2(\phi)=1-\phi^2/6\Mpl$. Therefore, the effective potential reduces to,
\bea
V^{\rm eff}=\frac{9}{4}\lambda\mpl^4\left(1-e^{-\sqrt{2/3}\frac{\phi}{\mpl}}\right)^2+\rho\sinh^2\left(\frac{\phi}{\sqrt{6}\mpl}\right).
\label{eq:veff}
\eea
We see that the new term behaves as $\sim\rho e^{\sqrt{2/3}\frac{\phi}{\mpl}}$ for positive $\phi$, and thus it will overturn the inflationary potential for $\phi/\mpl>\sqrt{3/2}$. Whether or not this affects inflation will also be dependent on the initial $\rho$, but we can in any case assume that $\rho_0\sim\lambda\mpl^4$ and continue our investigation. 
To analyse the model, we rescale the time by $t\rightarrow t/\sqrt{\lambda}$. Then (\ref{eq:veff}) becomes,
\bea
\varphi ''+3H\varphi'+3\sqrt{\frac{3}{2}}e^{-\sqrt{2/3}\varphi}\left(1-e^{-\sqrt{2/3}\varphi}\right)+\frac{\tilde{\rho}}{\sqrt{6}}\sinh\left(\sqrt{\frac{2}{3}}\varphi\right)=0,
\eea
with $\tilde{\rho}\equiv\frac{\rho}{\lambda}$. From now on we will call $\tilde{\rho}$ just $\rho$.
\begin{figure}[!ht]
\begin{center}
\includegraphics[scale=0.3]{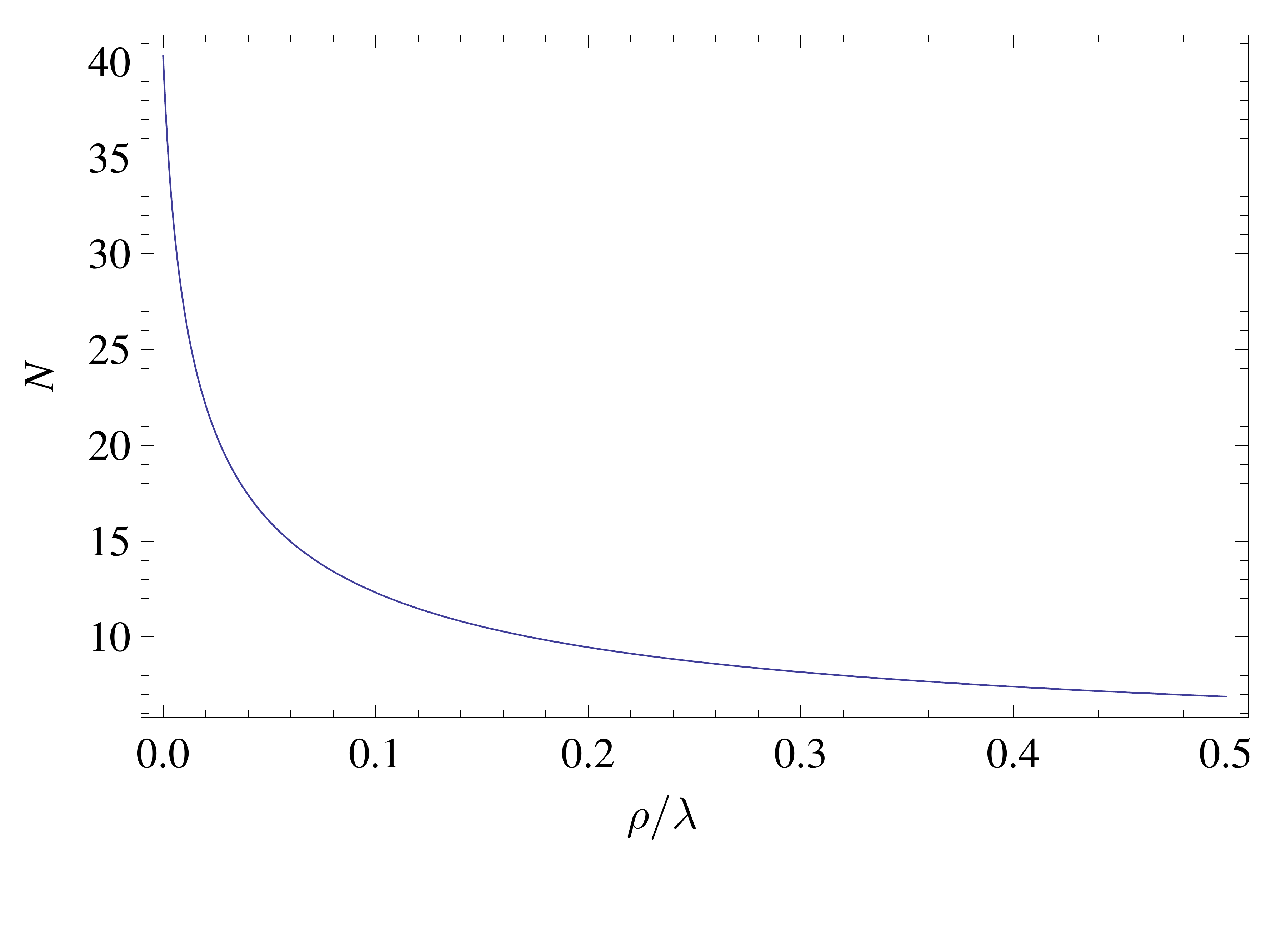}
\caption{Number of efolds before decaying into the attractor solution. Note that the scalar field decays rapidly even if  $ \rho  <1$. }
\label{plot-N}
\end{center}
\end{figure}
In the Starobinsky model, inflation occurs while the inflaton is rolling down the plateau  of the potential until it reaches the point  where $\epsilon$ becomes larger than 1. Once it enters this region the inflaton starts decaying until it reaches  the minimum of the potential where reheating starts. We can calculate the amount of e-folds that it would  take to cross this threshold  by using the slow roll approximation,
\bea
N\approx\int_{\varphi_I}^{\varphi_E}\frac{d\varphi}{\sqrt{2\epsilon_V}}.
\label{e-folds}
\eea
To simplify  this integral we can use  the equations of motion to rewrite $\sqrt{2\epsilon_V}$ and  then expand it in terms of $\rho$. Since we want to study how the matter is affected by the coupling we will assume that $\rho$ is constant. We then get that,

\bea
\frac{1}{\sqrt{2\epsilon_V}}\approx\frac{1}{\sqrt{2\epsilon_V}}_{\rho=0}-\frac{1}{\epsilon_V}_{\rho=0}\frac{e^{\sqrt{\frac{2}{3}}\varphi}}{9\sqrt{6}}\rho,
\eea
which is valid for small $\rho$ as long as the relation remains positive. Now we can see that  for small values of $\rho$ the integrand will decrease  when matter is added. This in turn will mean that the number of e-folds for which the inflaton remains in the inflationary sector of the potential severely decreases. 

Furthermore, we can integrate eq. (\ref{e-folds}) numerically, as shown in Fig.\ref{plot-N}. Indeed, we see that the field decays in a small number of e-folds and thus inflation is never realised  due to the steepness of the potential induced by the coupling term. 

We can be more systematic and generalise this result. To do so we note that the system has attractor solutions for fields starting to roll down from the left of the potential  due to the Hubble friction, irrespective of the value of $\rho$.

In the case of the Starobinsky model, fields falling into the attractor will keep rolling down producing an inflationary expansion until the slow roll condition no longer holds, when inflation terminates. Thus reviewing this attractor behaviour we can see what happens to the system when matter is included. 
Rewriting the Klein Gordon equation for the inflaton as follows,
\bea
\frac{d\dot\varphi}{d\varphi}&=&-\frac{3H\dot\varphi+V_{,\varphi}}{\dot\varphi}\\
&=&-\frac{3\dot\varphi\sqrt{\frac{\frac{\dot\varphi^2}{2}+V}{3}}+V_{,\varphi}}{\dot\varphi},
\label{eq:attractor}
\eea
we can analyse the phase space for the equations of motion. To find the attractor solutions  we assume that there is a trajectory for which  $\frac{d\dot\varphi}{d\varphi}\approx 0$. The solution for this equation is,
\bea
\dot\varphi^2=V\left(-1+\sqrt{1+\frac{(V_{,\varphi})^2}{3V^2}}\right),
\eea
which gives the behaviour of the kinetic energy in terms of the potential. 
Firstly consider the case when matter is absent.
In order to study inflationary trajectories in the Starobinsky model we assume that the field is far from the origin, thus $\phi>\sqrt{3/2}$.  Then from  (\ref{eq:attractor}), 
\bea
\dot\varphi^2=\frac{(V_{,\varphi})^2}{3V}=2e^{-2\sqrt{2/3}\varphi},
\eea 
where we see that when  $\phi$ is large the kinetic energy stays very  suppressed with respect to the potential, as expected by the slow roll conditions. However,  in the case when $\phi\sim\calO(1)$ the kinetic energy is the same order as the potential. When this happens the slow roll conditions is violated and inflation has ended. In the case when $\rho\gg 1$,  we have that $\frac{(V_{,\varphi})^2}{3V^2}\approx\frac{2}{9}\coth^2(\phi/\sqrt{6})$, inserting back into eq.~(\ref{eq:attractor}), we get that,
\bea
\dot\varphi^2\approx \frac{\sqrt{2}}{6}\rho\sinh(\sqrt{2/3}\varphi).
\eea
\begin{figure}[!ht]
\begin{center}
\includegraphics[scale=0.25]{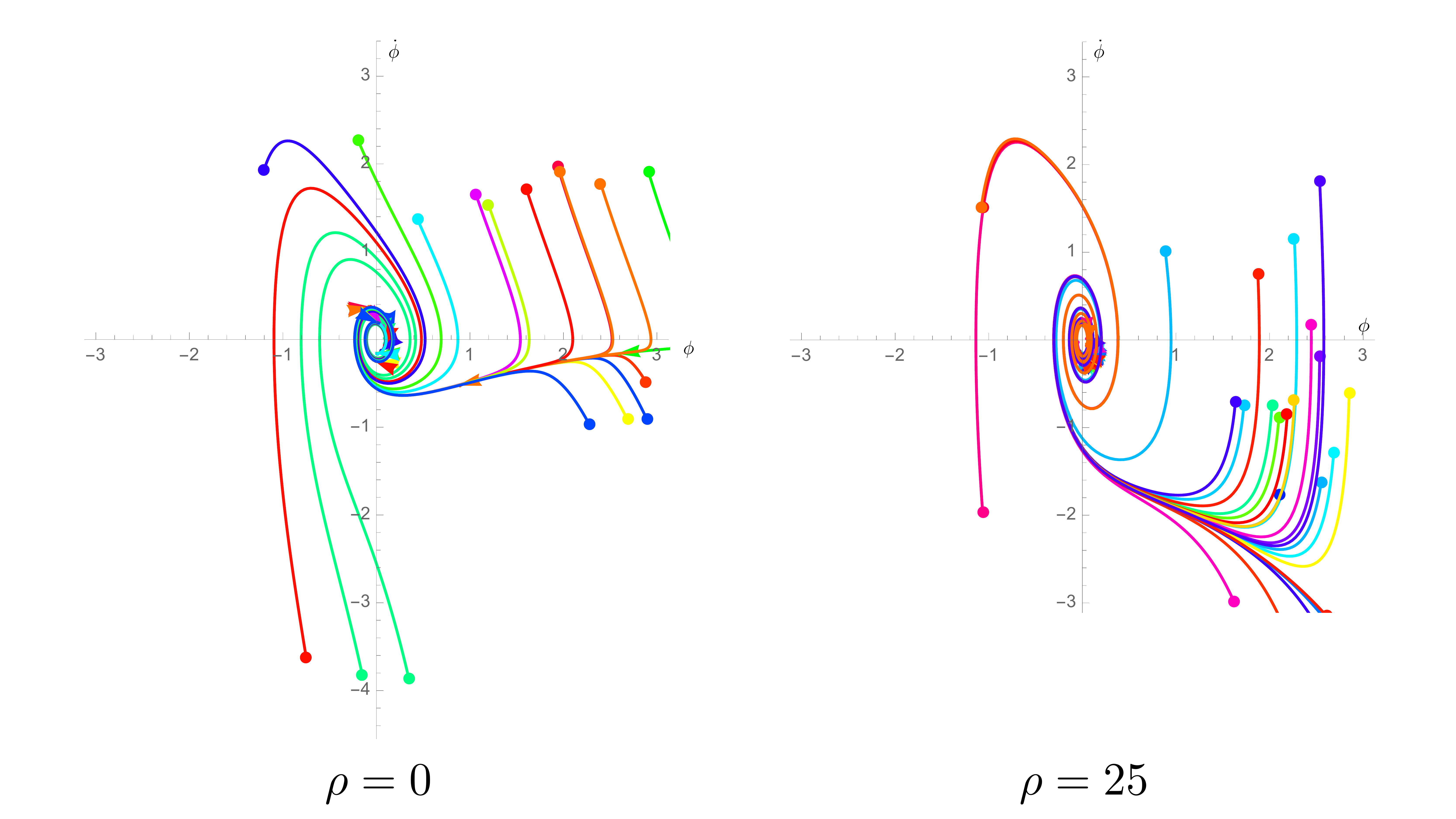}
\caption{Attractor behaviour for $\rho=0$ and $\rho=25$. Note that the kinetic energy is suppressed when $\rho=0$, but grows fast for $\rho=25$. Also is easy to notice that when matter is present the fields decays much faster out of the slow roll sector. }
\label{plot-Attractor}
\end{center}
\end{figure}
We now see that, as opposed to the previous case, the kinetic energy increases as we move away from the origin. Thus the slow roll conditions are no longer valid because the kinetic term is not negligible. Indeed, using this relation for $\dot\varphi^2$ we can calculate the Hubble ratio when $\rho$ is large,  
\bea
H^2= \frac{1}{3}\rho\sinh(\varphi/\sqrt{6})\left(\frac{\sqrt{2}}{3}\cosh(\varphi/\sqrt{6})+\sinh(\varphi/\sqrt{6})\right),
\eea
which is of the same order as the kinetic energy. This indicates that the effect of the coupling term is to uplift the kinetic energy in such a way that solutions decay to the origin exponentially faster, thus not allowing enough time for inflation to solve the horizon problem.   Indeed the Hubble friction will increase exponentially if we move away from the origin, thus the solution decays faster than in the case without matter. We plot the results in figure \ref{plot-Attractor}.
We can also calculate the time of decay. When $\rho= 0
$,
\bea
\varphi\propto-\sqrt{\frac{3}{2}}\log\left(\frac{2}{\sqrt{3}}t\right).
\eea
However, when $\rho\gg 1$,
\bea
\varphi\propto- 2 \sqrt{6} \tanh^{-1}(\tan(\frac{5t}{3 \sqrt{3}})).
\eea
We see that if there is sufficient matter, inflation will not happen because, for any suitable initial condition, the inflaton decays very fast to the bottom of the potential. As we show this result holds even if the matter density $\rho$ is of the same order of the energy density needed for inflation to start. We will now generalise this result to include other types of models.

\subsection{$\alpha$ model}
We start with the $\alpha$ model \cite{Kallosh:2013yoa} introduced in section (2). First recall that the scalar gravity action for this model,
\bea
S(\phi)= \int d^4x\sqrt{-g}\left\{\frac{1}{2}\left(1-\frac{\phi^2}{6} \right)^\alpha R-\frac{1}{2}\frac{\alpha-\alpha^2\phi^2/6}{(1-\phi^2/6)^{2-\alpha}} (\partial\phi)^2-V(\phi) \right\},\label{action:alphaJordan2}
\eea
where $\phi$ is the inflaton and we have set the rest of the fields to zero, fixed the conformal  symmetry by setting $X^6=\bar X^{\bar 6}=\sqrt{6}\mpl$. Also choosing $f(\phi/\sqrt{6})=  \frac{\lambda\phi^2}{(\phi/\sqrt{6}+1)^2}$  the potential in Einstein frame will be given by $V(\varphi)=V_0\left(1-e^{-\sqrt{\frac{2}{3\alpha}}\varphi}\right)^2$.
To study the phenomenology of the $\alpha$ models let us start by noting that for large $\alpha$  and for $\varphi\ll \sqrt{\frac{3\alpha}{2}}$, the potential term behaves as, 
\bea
V=\frac{m^2}{2}\phi^2,
\eea
where $m^2=\frac{4V_0}{3\alpha}$, whereas for small $\alpha$ it is the usual Starobinsky potential.  Therefore we can  divide our analysis in terms of the value of $\alpha$. Indeed, for $\alpha<\frac{\varphi^2}{\sqrt{6}}$, the inverse of the coupling function is $\Omega^{-1}\sim\sinh(\frac{\varphi}{\sqrt{6}\alpha})\ll1$. Therefore for large $\alpha$  we find that non-relativistic matter does not interfere and inflation may take place, provided that the initial conditions are set  not far from the origin. 

On the other hand for $\alpha\sim\calO(1)$, $\Omega^{-1}$ is large enough to interfere with the potential making it too steep for inflation to take place. Thus we have that as in the previous case, initial matter density can spoil inflation. Notice that this result is related to the curvature of the Kahler manifold, which is parametrized by $\alpha$, then it can be used to constrain possible scenarios. 
\subsection{D-term inflation}
D-term inflation~\cite{Binetruy:1996xj} is a supersymmetric completion of the hybrid inflation scenario. 
Whilst the orginal model suffers problems, the canonical superconformal model can be used to achieve a consistent  embedding~\cite{Buchmuller:2012ex}. 
For the large field  limit,  D-term inflation is equivalent to the Starobinsky model~\cite{Buchmuller:2013zfa}. Once the superconformal symmetry has been fixed the Kahler potential is given by,
\bea
\calK(z,\bar z)&=&3\ln\Omega(z,\bar z), \hspace{0.5cm}\mathrm{where}\\
\Omega^{-2}&=&1-\frac{1}{3}(\abs{S^2}+\abs{\phi_-}^2+\abs{\phi_+}^2)-\frac{\xi}{6}(S^2+\bar S^2),
\eea
with superpotential is $W=\lambda S\phi_+\phi_-$. The inflaton is $S$ and the other two fields are the waterfall fields. Notice that the  term proportional to $\xi$ breaks superconformal symmetry explicitly. For this model   the  potential becomes inflationary by radiative corrections. For an inflationary phase one has to consider a trajectory along the direction where just the real part of $S$ is non-zero. Then the non-canonical kinetic term in Einstein frame is given by 
\bea
K(\phi)=\frac{1}{1-\frac{1}{6}(1+\xi)\phi^2}\left(1+\frac{(1+\xi)\phi^2}{6(1-\frac{1}{6}(1+\xi)\phi^2)}\right).
\eea
The coupling function induces corrections to the potential, because canonically normalising the function is given by,
\bea
\Omega^2\sim1-\frac{(1-\xi)^2}{1+\xi}\tanh^2{\left(\frac{2\phi}{-\frac{\sqrt{6}}{\sqrt{1+\xi}}+\frac{\sqrt{6}\xi}.{\sqrt{1+\xi}}}\right)}
\eea
Thus we see that this function has the same shape as the one arising in conformal inflation, and therefore one has to expect that inflation cannot take place for a significant initial density of matter. 

\subsection{Universal attractors}
 The universal attractor  model studied  in ~\cite{Kallosh:2013hoa,Kallosh:2013maa} generalises  the models we have studied and therefore we can investigate deviations of the conformal case. Here, one has a non minimal coupling to gravity, and a potential given by 
 $\Omega=1+\xi f(\phi)$, $K=1$ and $U=\lambda f^2(\phi)\Omega^2$, where we used the notation from (\ref{action:NCInflation}). By varying the value of $\xi$ one can obtain different models of inflation. Thus for certain values the predictions will be equivalent to the ones of the Starobinsky model, whereas for others, to  chaotic inflation.

  This model can be brought into Einstein frame by a conformal transformation. We will start by considering the case when $\frac{\Omega'^2}{\Omega}\ll 1$. Then, the kinetic term reduces to,
  \bea
 S=\int d^4x\sqrt{-g} \left\{\frac{1}{2}R-\frac{1}{2\Omega}(\partial\phi)^2- V(\phi)\right\}.
\label{action:faf}
\eea
with corrections up to order $\calO(\epsilon^2)$ in slow roll parameters and $V=U/\Omega^2$. Let us specialise 
to the case $f(\phi)=\phi^2$. This case is equivalent to assuming that $\abs{\xi}\ll 1$. When integrating the kinetic term to obtain the 
canonically normalised field $\varphi$, we find that depending on the sign of $\xi$,
\bea
\phi&=&\frac{1}{\sqrt{\abs{\xi}}}\sinh(\sqrt{\abs{\xi}}\varphi)\hspace{1cm}\xi>0\\
\phi&=&\frac{1}{\sqrt{\abs{\xi}}}\sin(\sqrt{\abs{\xi}}\varphi)\hspace{1cm}\xi<0
\eea
which in turn leads to two similar situations. For positive $\xi$, we have a similar case to the one studied in \cite{Brax:2014baa} where the coupling function $\Omega$ will induce a change in the shape of the potential in such a way that the initial conditions for inflation are set. Indeed in this case, the potential $V\sim(\sinh(\sqrt{\xi}\varphi))^4$ has a minima at $\varphi=0$ which is uplifted when there is an initial density of matter by a term $\sim\rho\cosh^{-2}({\sqrt{\xi}\phi})$. Then, the inflaton will decay to a new minima  until  matter flushes away and inflation starts. 

On the other hand for negative and small  $\xi$ the case is analogous to the one studied at the beginning of this section. However it has some variations, here the potential will be $V\sim \sin^4(\sqrt{\xi}\varphi)$, which is similar to  a chaotic potential for $\abs{\varphi} <\pi/(2\sqrt{\xi})$. Indeed, the potential has a plateau for small values of $\varphi$ where inflation can occur. In the presence  of matter the dominant term will be  $\sim \rho\cos^{-2}(\sqrt{\xi}\varphi)$. This term grows faster than the original potential and will uplift the plateau,  thus the inflaton will decay fast to the origin, similar to the case for the conformal coupling $\chi=-1/6$.

 For the strong coupling limit $\xi\gg1$ ~\cite{Kallosh:2013tua}, we have that  $\frac{\Omega'^2}{\Omega}\gg \frac{1}{\Omega^2}$, and then 
 \bea
 S=\int d^4x\sqrt{-g} \left\{\frac{1}{2}R-\frac{3\Omega'^2}{2\Omega^2}(\partial\phi)^2- V(\phi)\right\}, 
\label{action:faf}
 \eea
Normalising the kinetic term  the new variable is $\varphi=\pm\sqrt{\frac{3}{2}}\log \Omega(\phi)$. Choosing the correct sign to obtain an inflationary potential one finds $V(\varphi)=\frac{\lambda}{\xi^2}\left(1-e^{-\sqrt{2/3}\varphi}\right)^2$, is independent of the choice of  $f(\phi)$. The potential now resembles the Starobinsky model. Note also that this result is independent of the sign of $\xi$.  Moreover, the coupling to gravity is, 
\bea
\Omega(\varphi)=e^{\sqrt{2/3}\varphi},
\eea
and then, the coupling to matter  decays exponentially for large $\phi$. However,  for large non-relativistic matter $\rho$ the potential will not  have a local minima because the function $\Omega(\phi)^{-1}$ grows for negative values. Hence, the original minimum of the Starobinsky potential is 
no longer a minimum in this case. We see that this is rather different to 
the other cases we have considered and seems to be a generic feature for this kind of situation. To further proceed one needs to study the reheating process for this model, and how the change in the minima will affect it. 
\begin{figure}[!ht]
\begin{center}
\includegraphics[scale=0.25]{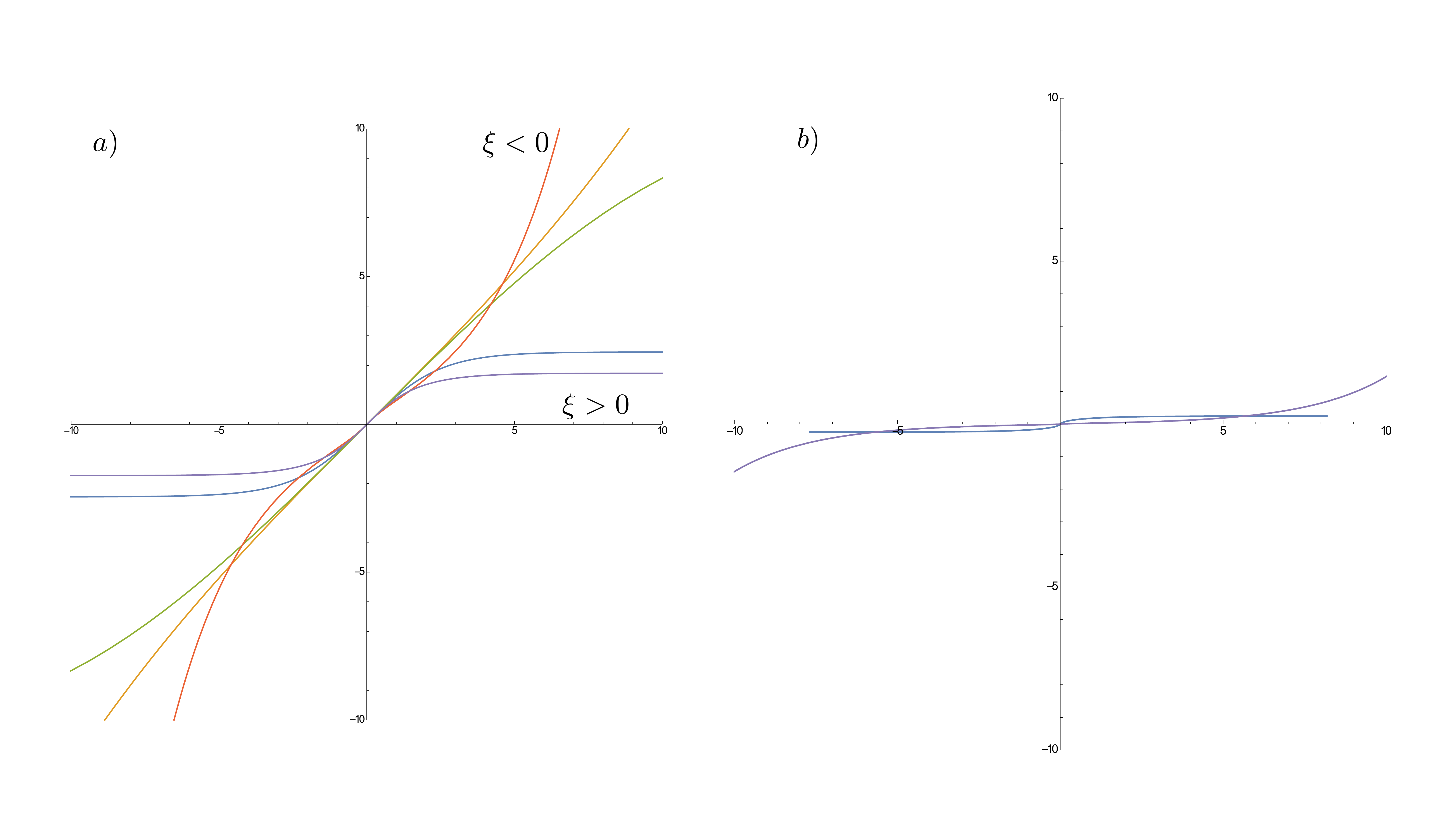}
\caption{a)Plot of $\varphi$ for various values of $\xi$. Note that there is a change in the curvature depending on the sign of $\xi$. In b) the plot shows the canonically normalise fields for large values of $\xi$ of both signs, and it is noticeable that both are similar in the region where inflation happens.  }
\label{plot-canfields}
\end{center}
\end{figure} 

 To examine the intermediate case, we need to use numerical integration. To do so, let us assume again that,
\bea
\Omega(\phi)=1+\xi\phi^2.
\eea
Now, the canonical variable will be given by,
\bea
\frac{d \varphi}{d\phi}=\frac{\sqrt{1+\xi \phi^2+6\xi^2\phi^2}}{1+\xi \phi^2}.
\label{eq:canonic}
\eea
First note that for small $\xi$ the canonical term is approximately flat and the potential is quartic for both signs. Therefore inflation occurs in the flat section near the origin. However, we see from (\ref{eq:canonic}) that the result of this integral will change, depending on the sign of $\xi$ and thus the coupling to matter will also change. This gives the different situations  
described previously. To see this more clearly, we plot the values of $\xi$ in Fig. \ref{plot-canfields}. We see that when $\xi<0$ the slope changes and the coupling function will be similar to a smoothed $\cos^{-2}\varphi$, whereas for the other case will be a $\cosh^{-2}\varphi$. On the contrary, when $\xi$ is large near the origin the potential will be too steep and  inflation occurs in the plateau characteristic, of the Starobinsky model.
In the intermediate situation when $\xi$ is increased the potential near the origin will become  steep, but the potential will change to $V(\varphi)\sim\left(1-e^{-\sqrt{2/3}\varphi}\right)^2$, and then inflation will occur in the plateu of this potential. Increasing $\xi$ will diminish the effect of the coupled matter, because the coupling will behave as $e^{-\sqrt{2/3}\varphi}$ and then for $\phi\sim\calO(1)$ will be supressed. 

Finally, let us again note , that it is not sufficient to have negative $\xi$, to affect inflation. This is because for very large $\xi$, there is a strong coupling regime in which the overall form of $\Omega$  and therefore the sign of $\xi$ is not important. Indeed, we see from the equation for (\ref{eq:canonic}) that for large $\phi$ the result is independent of $\xi$, and thus the inverse of the coupling function will behave exactly as we discussed earlier.


\section{Discussion}
\setcounter{equation}{0}

We have shown that the presence of non-relativistic matter can have a noticeable affect on  non-canonical inflationary theories. This is due to the fact that when coupled matter is taken into consideration it can change the shape of the inflationary potential for enough time to affect its dynamics.  We focused on the particular case of conformal models which are equivalent to the Starobinsky model. These models are interesting because the initial action in Jordan frame leads to a unique coupling to matter when transformed to Einstein  frame. In this case the resulting potential is  too steep for inflation to begin when there is an initial density of matter present.  We proved that although an initial density of matter will  decay rapidly due to cosmic expansion,  coupled matter changed the attractor behaviour of the theory and that generally the inflaton  will decay to the origin in a few efolds. This result is  consistent with  \cite{Brax:2014baa,Dong:2013swa}, where it was shown that an initial matter density can set the initial conditions before decaying.   

We first considered the Starobinsky model. In this case we can study the phenomenology in detail. Here since inflation can only take place in a plateau away of the origin, the field cannot start to produce inflation once matter is coupled because the potential will be too steep.  We showed that this situation arises naturally in the superconformal supergravity theories of inflation because there is an extra symmetry required that fixes the coupling to gravity. We showed that in the case of an SO(1,1) symmetry the coupling will be induced by the symmetry. Using this, we generalised to include other models. Thus in this case the coupling is a feature of the symmetry rather than the particular model. We showed that in more general models, such as $\alpha$-models  there is a similar behaviour, although this case depends on the geometry of the Kahler manifold.

Moreover, as reported previously \cite{Brax:2014baa}, the inclusion of coupled matter can also set the initial conditions for inflation. The difference arises from the geometry of the field space, or  in supergravity from the Kahler manifold. We prove that by changing the curvature parameter in the universal attractror one can either  spoil inflation or naturally set its initial conditions. It is particulary interesting that the more symmetric cases were the ones with more dramatic consequences. Indeed it is very easy to see that the result we obtain for the conformally invariant model can be generalised for the $SO(1,N)$ case.

Our results hold independently of the frame chosen. For example, if one  gauges the theory by choosing $\chi^2-\phi^2=\sqrt{6}$ on (\ref{action:conformalinv}), then one is taken directly to the Einstein frame  of the theory (\ref{action:inflStarobinsky}) with coupling function $\Omega$ is equal to $1$. Because of this, one may think that there is a possible ambiguity between both frames,  but, this is no longer true if the situation is considered more carefully. Taking a step back one  can perform a full gauge invariant calculation to show that the equation of motion are,
\bea
\Box\varphi-V_{,\varphi}-\Omega_{,\varphi}\Omega^3T^J=0,
\label{eq:gaugeinv}
\eea
where $ \Omega(\varphi)=\Omega(\varphi(\phi^2-\chi^2))$, ie, in general $\Omega$ is a function of both fields and it should not be removed by a gauge choice. Then one can see, our results holds in any gauge. 
Furthermore, the result in ~\cite{Domenech:2015qoa} where a massless curvaton field coupled  to inflation was considered, showed physical observables were the same, though there was a different  physical interpretation in both frames.
Notice also that this result is  different to the $\eta$ problem, where inflation is unviable because the inflaton mass receives large radiative correction. Indeed for the particular supergravity studied  a shift symmetry  avoids the corrections at all energies and therefore we can isolate this effect from the  inclusion of non-relativistic matter. In this sense  one can estimate the value of $\rho$ for which inflation is affected. We showed in the Starobinsky model that   for $\rho\sim \lambda\mpl^4$ the potential was too steep. Since  inflation occurs  at energies $\frac{V}{\mpl^4}\ll\Delta_\calR\sim 10^{9}$, and therefore $\lambda\sim 10^{-9}$, we have that $\rho\ll\mpl^9$, and the theory is stable at this level. 
We also proved that this effect arises in the more general universal attractor models. Here it is easy to appreciate how the geometry of the target space 
affects the coupling to matter in such a way that for certain cases inflation cannot take place, while for other cases works as a mechanism to set the initial 
conditions. 
The initial density of non-relativistic matter could arise as a tower of Kaluza Klein modes, and thus it will appear in models with extra dimensions. Whilst we have not analysed the specific detail that coupled matter could have, one would have expected the effects to be negligible due to the cosmic expansion.  We have shown otherwise. An embedding of the initial matter in a particular UV model is still an interesting possibility for further research.

\section{Acknowledgments}
We would like to thank Raquel Ribeiro for useful discussions. 
This work is funded in part by STFC under grants ST/L000385/1 and ST/L000636/1. SC is funded by Conicyt through  Becas Chile  and the Cambridge Commonwealth, European and International Trust. 

\providecommand{\href}[2]{#2}\begingroup\raggedright\endgroup

\begin{thebibliography}{99}


\bibitem{Benson:2014qhw}
  B.~A.~Benson {\it et al.}  [SPT-3G Collaboration],
  Proc.\ SPIE Int.\ Soc.\ Opt.\ Eng.\  {\bf 9153} (2014) 91531P
  [arXiv:1407.2973 [astro-ph.IM]].
\bibitem{Keck}
  [Keck Array Collaboration],
\bibitem{Alvarez:2014vva}
  M.~Alvarez, T.~Baldauf, J.~R.~Bond, N.~Dalal, R.~de Putter, O.~Doré, D.~Green and C.~Hirata {\it et al.},
  arXiv:1412.4671 [astro-ph.CO].
  \bibitem{Bezrukov:2007ep}
F.~L. Bezrukov and M.~Shaposhnikov, {\it {The Standard Model Higgs boson as the
  inflaton}},  {\em Phys.Lett.} {\bf B659} (2008) 703--706,
  [\href{http://xxx.lanl.gov/abs/0710.3755}{{\tt arXiv:0710.3755}}].
\bibitem{Cheung:2007st}
  C.~Cheung, P.~Creminelli, A.~L.~Fitzpatrick, J.~Kaplan and L.~Senatore,
  JHEP {\bf 0803} (2008) 014
  [arXiv:0709.0293 [hep-th]].
  


\bibitem{Ade:2015lrj}
  P.~A.~R.~Ade {\it et al.}  [Planck Collaboration],
  arXiv:1502.02114 [astro-ph.CO].
\bibitem{Martin:2013tda}
J.~Martin, C.~Ringeval, and V.~Vennin, {\it {Encyclopaedia Inflationaris}},
  \href{http://xxx.lanl.gov/abs/1303.3787}{{\tt arXiv:1303.3787}}.

\bibitem{Baumann:2014nda}
  D.~Baumann and L.~McAllister,
  arXiv:1404.2601 [hep-th].
  
\bibitem{Baumann:2011su}
  D.~Baumann and D.~Green,
  JCAP {\bf 1109} (2011) 014
  [arXiv:1102.5343 [hep-th]].
  
\bibitem{Achucarro:2010da}
  A.~Achucarro, J.~O.~Gong, S.~Hardeman, G.~A.~Palma and S.~P.~Patil,
  JCAP {\bf 1101} (2011) 030
  [arXiv:1010.3693 [hep-ph]].
  
\bibitem{Avgoustidis:2012yc} 
  A.~Avgoustidis, S.~Cremonini, A.~C.~Davis, R.~H.~Ribeiro, K.~Turzynski and S.~Watson,
  JCAP {\bf 1206}, 025 (2012)
  [arXiv:1203.0016 [hep-th]].
\bibitem{Achucarro:2012yr}
  A.~Achucarro, V.~Atal, S.~Cespedes, J.~O.~Gong, G.~A.~Palma and S.~P.~Patil,
  Phys.\ Rev.\ D {\bf 86} (2012) 121301
  [arXiv:1205.0710 [hep-th]].
 
 
\bibitem{Baumann:2011nk}
  D.~Baumann and D.~Green,
  Phys.\ Rev.\ D {\bf 85} (2012) 103520
  [arXiv:1109.0292 [hep-th]].
  
\bibitem{Lyth:1998xn}
  D.~H.~Lyth and A.~Riotto,
  Phys.\ Rept.\  {\bf 314} (1999) 1
  [hep-ph/9807278].
  \bibitem{Damour:2002mi}
T.~Damour, F.~Piazza, and G.~Veneziano, {\it {Runaway dilaton and equivalence
  principle violations}},  {\em Phys.Rev.Lett.} {\bf 89} (2002) 081601,
  [\href{http://xxx.lanl.gov/abs/gr-qc/0204094}{{\tt gr-qc/0204094}}].
  
\bibitem{Starobinsky:1980te}
  A.~A.~Starobinsky,
  Phys.\ Lett.\ B {\bf 91} (1980) 99.
  
\bibitem{Whitt:1984pd} 
  B.~Whitt,
  Phys.\ Lett.\ B {\bf 145}, 176 (1984).
\bibitem{Cecotti:1987sa}
  S.~Cecotti,
  Phys.\ Lett.\ B {\bf 190} (1987) 86.
  
\bibitem{Cecotti:1987qe}
  S.~Cecotti, S.~Ferrara, M.~Porrati and S.~Sabharwal,
  Nucl.\ Phys.\ B {\bf 306} (1988) 160.
  
  
\bibitem{Ketov:2010qz}
  S.~V.~Ketov and A.~A.~Starobinsky,
  Phys.\ Rev.\ D {\bf 83} (2011) 063512
  [arXiv:1011.0240 [hep-th]].
  \bibitem{Ellis:2013xoa}
J.~Ellis, D.~V. Nanopoulos, and K.~A. Olive, {\it {No-Scale Supergravity
  Realization of the Starobinsky Model of Inflation}},  {\em Phys.Rev.Lett.}
  {\bf 111} (2013) 111301, [\href{http://xxx.lanl.gov/abs/1305.1247}{{\tt
  arXiv:1305.1247}}].
  \bibitem{Kallosh:2013lkr}
R.~Kallosh and A.~Linde, {\it {Superconformal generalizations of the
  Starobinsky model}},  {\em JCAP} {\bf 1306} (2013) 028,
  [\href{http://xxx.lanl.gov/abs/1306.3214}{{\tt arXiv:1306.3214}}].
  
  \bibitem{Buchmuller:2013zfa}
W.~Buchmuller, V.~Domcke, and K.~Kamada, {\it {The Starobinsky Model from
  Superconformal D-Term Inflation}},  {\em Phys.Lett.} {\bf B726} (2013)
  467--470, [\href{http://xxx.lanl.gov/abs/1306.3471}{{\tt arXiv:1306.3471}}].

\bibitem{Farakos:2013cqa}
  F.~Farakos, A.~Kehagias and A.~Riotto,
  Nucl.\ Phys.\ B {\bf 876} (2013) 187
  [arXiv:1307.1137].
  \bibitem{Ellis:2013nxa}
J.~Ellis, D.~V. Nanopoulos, and K.~A. Olive, {\it {Starobinsky-like
  Inflationary Models as Avatars of No-Scale Supergravity}},  {\em JCAP} {\bf
  1310} (2013) 009, [\href{http://xxx.lanl.gov/abs/1307.3537}{{\tt
  arXiv:1307.3537}}].

\bibitem{Kallosh:2013maa}
  R.~Kallosh and A.~Linde,
  JCAP {\bf 1310} (2013) 033
  [arXiv:1307.7938].

\bibitem{Kallosh:2013daa}
R.~Kallosh and A.~Linde, {\it {Multi-field Conformal Cosmological Attractors}},
   \href{http://xxx.lanl.gov/abs/1309.2015}{{\tt arXiv:1309.2015}}.
\bibitem{Kallosh:2013tua}
  R.~Kallosh, A.~Linde and D.~Roest,
  Phys.\ Rev.\ Lett.\  {\bf 112} (2014) 1,  011303
  [arXiv:1310.3950 [hep-th]].
\bibitem{Mosk:2014cba}
  B.~Mosk and J.~P.~van der Schaar,
  JCAP {\bf 1412} (2014) 12,  022
  [arXiv:1407.4686 [hep-th]].

\bibitem{Freedman:2012zz}
  D.~Z.~Freedman and A.~Van Proeyen,
  Cambridge, UK: Cambridge Univ. Pr. (2012) 607 p
\bibitem{Buttazzo:2013uya}
D.~Buttazzo, G.~Degrassi, P.~P. Giardino, G.~F. Giudice, F.~Sala, et~al., {\it
  {Investigating the near-criticality of the Higgs boson}},
  \href{http://xxx.lanl.gov/abs/1307.3536}{{\tt arXiv:1307.3536}}.
  
 \bibitem{Ferrara:2010yw}
S.~Ferrara, R.~Kallosh, A.~Linde, A.~Marrani, and A.~Van~Proeyen, {\it {Jordan
  Frame Supergravity and Inflation in NMSSM}},  {\em Phys.Rev.} {\bf D82}
  (2010) 045003, [\href{http://xxx.lanl.gov/abs/1004.0712}{{\tt
  arXiv:1004.0712}}].

\bibitem{Ferrara:2010in}
S.~Ferrara, R.~Kallosh, A.~Linde, A.~Marrani, and A.~Van~Proeyen, {\it
  {Superconformal Symmetry, NMSSM, and Inflation}},  {\em Phys.Rev.} {\bf D83}
  (2011) 025008, [\href{http://xxx.lanl.gov/abs/1008.2942}{{\tt
  arXiv:1008.2942}}].
  
\bibitem{deRham:2014wfa} 
  C.~de Rham and R.~H.~Ribeiro,
  JCAP {\bf 1411}, no. 11, 016 (2014)
  [arXiv:1405.5213 [hep-th]].
\bibitem{Joyce:2014kja}
  A.~Joyce, B.~Jain, J.~Khoury and M.~Trodden,
  Phys.\ Rept.\  {\bf 568} (2015) 1
  [arXiv:1407.0059 [astro-ph.CO]].

\bibitem{Hinterbichler:2010es}
K.~Hinterbichler and J.~Khoury, {\it {Symmetron Fields: Screening Long-Range
  Forces Through Local Symmetry Restoration}},  {\em Phys.Rev.Lett.} {\bf 104}
  (2010) 231301, [\href{http://xxx.lanl.gov/abs/1001.4525}{{\tt
  arXiv:1001.4525}}].



\bibitem{Brax:2014baa}
  P.~Brax and A.~C.~Davis,
  JCAP {\bf 1405} (2014) 019
  [arXiv:1401.7281 [astro-ph.CO]].

\bibitem{Damour:1994zq}
T.~Damour and A.~M. Polyakov, {\it {The String dilaton and a least coupling
  principle}},  {\em Nucl.Phys.} {\bf B423} (1994) 532--558,
  [\href{http://xxx.lanl.gov/abs/hep-th/9401069}{{\tt hep-th/9401069}}].



\bibitem{Kallosh:2013yoa}
R.~Kallosh, A.~Linde, and D.~Roest, {\it {Superconformal Inflationary
  $\alpha$-Attractors}},  {\em JHEP} {\bf 1311} (2013) 198,
  [\href{http://xxx.lanl.gov/abs/1311.0472}{{\tt arXiv:1311.0472}}].


\bibitem{Kehagias:2013mya}
  A.~Kehagias, A.~M.~Dizgah and A.~Riotto,
  Phys.\ Rev.\ D {\bf 89} (2014) 4,  043527
  [arXiv:1312.1155 [hep-th]].


\bibitem{Kallosh:2013hoa}
R.~Kallosh and A.~Linde, {\it {Universality Class in Conformal Inflation}},
  {\em JCAP} {\bf 1307} (2013) 002,
  [\href{http://xxx.lanl.gov/abs/1306.5220}{{\tt arXiv:1306.5220}}].

  
\bibitem{Kallosh:2010xz}
  R.~Kallosh, A.~Linde and T.~Rube,
  Phys.\ Rev.\ D {\bf 83} (2011) 043507
  [arXiv:1011.5945 [hep-th]].
\bibitem{Ferrara:2013rsa}
  S.~Ferrara, R.~Kallosh, A.~Linde and M.~Porrati,
  Phys.\ Rev.\ D {\bf 88} (2013) 8,  085038
  [arXiv:1307.7696 [hep-th]].

\bibitem{Buchmuller:2012ex}
  W.~Buchm�ller, V.~Domcke and K.~Schmitz,
  JCAP {\bf 1304} (2013) 019
  [arXiv:1210.4105 [hep-ph]].
  
\bibitem{Hertzberg:2014aha}
  M.~P.~Hertzberg,
  Phys.\ Lett.\ B {\bf 745} (2015) 118
  [arXiv:1403.5253 [hep-th]].
  
\bibitem{Jackiw:2014koa}
  R.~Jackiw and S.~Y.~Pi,
  Phys.\ Rev.\ D {\bf 91} (2015) 6,  067501
  [arXiv:1407.8545 [gr-qc]].
  \bibitem{Kallosh:2013pby}
R.~Kallosh and A.~Linde, {\it {Superconformal generalization of the chaotic
  inflation model $\frac{\lambda}{4} \phi^{4} - \frac{\xi}{2} \phi^{2}R$}},
  {\em JCAP} {\bf 1306} (2013) 027,
  [\href{http://xxx.lanl.gov/abs/1306.3211}{{\tt arXiv:1306.3211}}].
  
\bibitem{Dong:2013swa} 
  R.~Dong, W.~H.~Kinney and D.~Stojkovic,
  JCAP {\bf 1401}, no. 01, 021 (2014)
  [arXiv:1307.4451 [astro-ph.CO]].
\bibitem{GarciaBellido:2011de}
J.~Garcia-Bellido, J.~Rubio, M.~Shaposhnikov, and D.~Zenhausern, {\it
  {Higgs-Dilaton Cosmology: From the Early to the Late Universe}},  {\em
  Phys.Rev.} {\bf D84} (2011) 123504,
  [\href{http://xxx.lanl.gov/abs/1107.2163}{{\tt arXiv:1107.2163}}].
  
\bibitem{Binetruy:1996xj}
  P.~Binetruy and G.~R.~Dvali,
  Phys.\ Lett.\ B {\bf 388} (1996) 241
  [hep-ph/9606342].
  
%
%

\bibitem{Domenech:2015qoa}
  G.~Dom�nech and M.~Sasaki,
  arXiv:1501.07699 [gr-qc].


%
%
%
%
%
%
%
%
%
%
%
%
%
%
%
%
%
%
%
%
%
%
%
%
%
%
%
%
%

\end{thebibliography}
\end{document}